\documentclass[preprint,preprintnumbers,amsmath,amssymb,prd,tightenlines
]{revtex4-1}
\usepackage{graphicx}
\begin{document}
\title{Revision of the brick wall method for calculating the black hole thermodynamic quantities}
\newpage
\author{F. Lenz $^{{\rm a,c}}$}    \email{flenz@theorie3.physik.uni-erlangen.de} 
\author{K. Ohta $^{{\rm b}}$}      \email{ohta@nt1.c.u-tokyo.ac.jp}
\author{K. Yazaki $^{{\rm c}}$}  \email{yazaki@phys.s.u-tokyo.ac.jp}
\affiliation{$^{{\rm a}}$ Institute for Theoretical Physics III \\
University of Erlangen-N\"urnberg \\
Staudtstrasse 7, 91058 Erlangen, Germany\\ \\
$^{{\rm b}}$ Institute of Physics \\  University of Tokyo\\ Komaba, Tokyo 153-8902, Japan  \\ \\
$^{{\rm c}}$ Hatsuda Quantum Hadron Physics Laboratory\\ Nishina Center, RIKEN\\
Wako, Saitama 351-0198, Japan \\} 
\date{September 1, 2015} 
\begin{abstract}
Within the framework of the  ``brick wall model'', a novel method is developed to compute the contributions of a scalar field to the thermodynamic quantities of black holes. The relations between   (transverse) momenta and frequencies in Rindler space are determined numerically with high accuracy and analytically with an accuracy of  better than 10\,\% and are compared with the corresponding quantities in Minkowski space. In conflict with earlier results, the thermodynamic properties of black holes turn out  to be those of a low temperature system. The resulting discrepancy for partition function and entropy by two orders of magnitude  is analyzed in detail. 
In the final part we carry out the analogous studies for scalar fields in de Sitter space and thereby confirm that our method applies also to the important case of spherically symmetric spaces.    
\end{abstract}
\maketitle    
\section{Introduction} 
The relations between momenta and energies are fundamental quantities in statistical physics. They determine the density of states and therefore the partition function and other thermodynamic quantities.  Similarly,  the momentum-frequency relations (``m-f relations'')   of fields in the presence of a Killing  horizon determine the corresponding thermodynamic quantities. Before calculating them, singularities of the fields at the horizon have to be regularized. We adopt the most commonly used ``brick wall'' method \cite{tHOO85,SUUG94,tHOO96} which regularizes  these singularities by restricting the fields  to a region close to  but  outside the horizon. Within this framework  a variety of investigations (cf.\,\cite{MUIS98} or the reviews\,\cite{FRFU98},\,\cite{PADM05} where also other approaches are discussed) have been carried out and  corrections to the semiclassical approximation have been
investigated, cf.\,\cite{SSSS08},\,\cite{KIKU13}. 
Other methods have been applied and related  to  the brick wall method such as the regularization via a Pauli-Villars method \cite{DELM95}. 
In the most commonly applied procedure, the density of states of massless fields is  evaluated  in a modified   WKB approximation where  in addition it is assumed that the summation  over the modes can be replaced by an integration. To the best of our knowledge, the validity of these approximations has never been verified. It will be analyzed in detail.     

We will  apply the brick wall model for scalar fields in  Rindler space-time. However we will neither assume  a priori that the WKB approximation is appropriate nor  that the discrete set of eigenvalues   can be replaced  by a continuum. 
We will present ``exact'' numerical results for the m-f relations and the horizon induced   partition function and  entropy. To provide insight we also   will present various approximate, analytical results which together cover the whole range of frequencies.   We will show that in the WKB approximation  the numerically determined value of the partition function is underestimated by  a factor of 3. It is  overestimated by two orders of magnitude 
 if the sum over the discrete modes  is replaced by an integral which can be determined analytically.  It will be seen that the partition function is, up to 1\,\%, given by the lowest frequency mode. The source of this large discrepancy will be identified and its consequences will be discussed.  We also will
show that finite mass effects are visible only under extreme conditions.  

In order to demonstrate the validity of our method applied to  spherically symmetric spaces we have chosen to calculate the thermodynamic quantities of  scalar fields in  de Sitter space  (in the static metric) where not only numerical but again also approximate analytical results can be obtained.   We will establish the connection  between the angular momentum-frequency relations (``am-f relations'')  of  de Sitter space and the m-f relations  of Rindler space  cf.\,\cite{PADM86}.  
\section{Momentum-frequency relations  of  scalar fields in Rindler  space}\label{dprri}
{\bf Scalar Fields in Rindler Space}
\vskip .1cm
A uniformly accelerated observer in  Minkowski space moves along the hyperbola \cite{RIND01}
\begin{equation*}
x^{2} - t^{2} = \frac{1}{a^{2}}\,,\quad  
{\bf x}_{\perp}= 0\,, 
\label{hyper}
\end{equation*}
with the acceleration  denoted by $a$ and the coordinates  transverse to the motion by  ${\bf x}_{\perp}$. After the coordinate transformation 
\begin{equation}t,x,{\bf x}_{\perp}\to \tau,\xi,{\bf x}_{\perp}:\quad t (\tau, \xi) = \frac{1}{a} \; e^{a \xi} \sinh a \tau\,, \quad 
x (\tau, \xi) = \frac{1}{a} \; e^{a \xi} \cosh a \tau \,,
\label{rimi}
\end{equation}
a particle at rest in the observer's system at $\xi=\xi_0$  corresponds to the uniformly accelerated motion  in Minkowski space with acceleration $ae^{-a\xi_0}$.  
The space-time defined by Eq.\,(\ref{rimi}) is  the  Rindler space with the metric
\begin{equation}
ds^{2} = e^{2 \kappa \xi} (d \tau^{2} - d \xi^{2})- d{\bf x}^2_{\perp} \,,  \quad \kappa=a\,,           
\label{rin}
\end{equation}
where the identity of the acceleration $a$  and  the surface gravity $ \kappa$ has been used,  cf.\,\cite{WALD94}.
The coordinate transformation  (\ref{rimi}) is not one-to-one. The coordinates $-\infty < \tau, \xi < \infty $ cover only one quarter of the Minkowski space, the  ``Rindler wedge''$R_+$ 
\begin{equation*}
R_{\pm} = \big\{x^\mu\big|\,|t|\le \pm x \big\}\,.                          
\label{rau3}
\end{equation*}
Upon reversion of the sign of $x$ in Eq.\,(\ref{rimi}) it is the  Rindler wedge $R_-$ which is covered by the corresponding parametrization. No causal connection exists between the two Rindler wedges $R_\pm$.

\begin{figure}[h]
\begin{center}
\includegraphics[width=.6\linewidth]{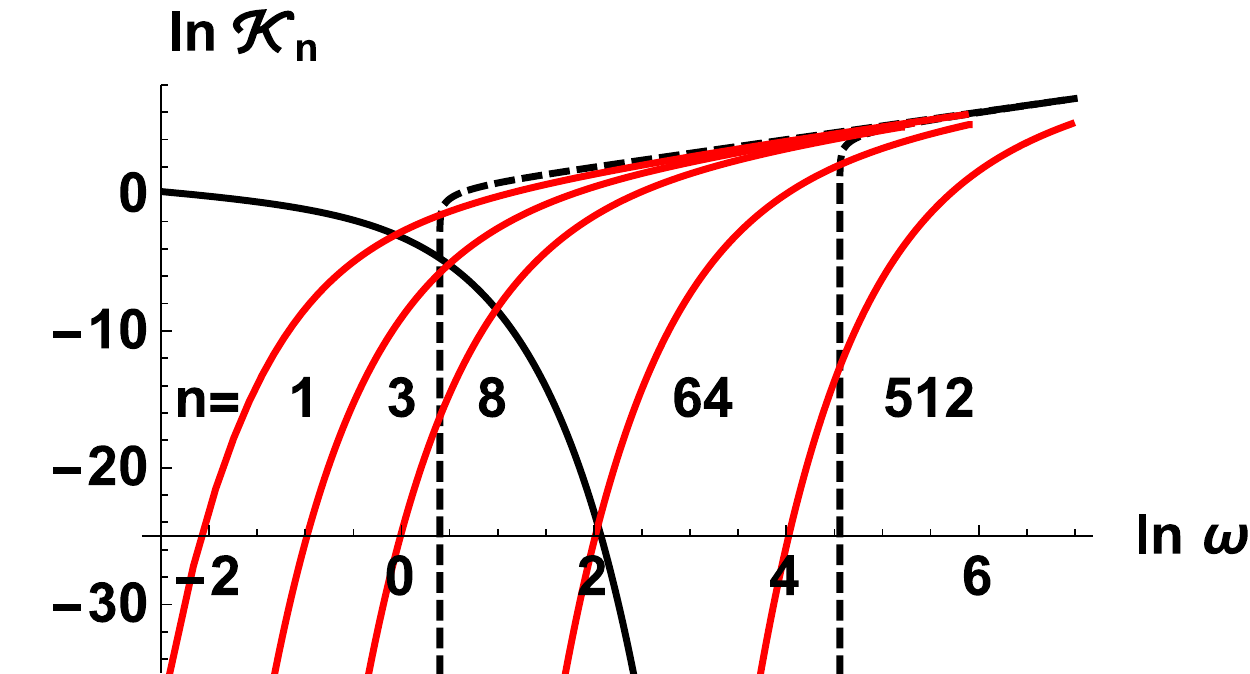}
\caption{Double logarithmic plots of  numerically determined momentum-frequency relations  ${\mathcal K}_n(\omega)$ (Eq.\,(\ref{bocu})) in Rindler space for five  values of $n$ (solid, red curves) are compared with m-f relations (Eq.(\ref{dispm1}))  in Minkowski space (dashed, black curves) with  $\lambda=2.1$ and  $n=1$ and $64$. The curve decreasing with $\ln \omega$ is half the  logarithm of the ``Boltzmann factor'' $-0.5\, \ln (e^{\beta\omega}-1),$ cf.\,Eq.\,(\ref{zet2}) (solid black curve).}
\vskip-.4cm
\label{disprel}
\end{center}
\end{figure}

 We consider a  non-interacting scalar field $\phi$ in   Rindler space  with the action
\begin{equation}
S = \frac{1}{2} \int d \tau \, d \xi \,d{\bf x}_{\perp} \big \{ (\partial_{\tau} \phi )^{2}
- (\partial_{\xi} \phi )^{2} - (m^{2} \phi^{2} + (\boldsymbol{\nabla}_{\perp} \phi)^{2} \big) \;
 e^{2\kappa \xi} \big \} \,,                                                     \label{ac}
\end{equation}
which is nothing else than the Minkowski space action restricted to one of its quarters. The solutions of the equations of motion, vanishing exponentially with  $\xi\to\infty$,   read
\begin{equation}
\phi(\tau,\xi,{\bf x}_\perp)= e^{- i \omega \tau}e^{i {\bf k} _{\perp} {\bf x}_{\perp}}\,K_{i\omega/\kappa} \big(z(\xi)\big)\,,
 \quad z(\xi)=\frac{1}{\kappa}\sqrt{m^2+k_\perp^2}\,e^{\kappa\xi}\,,
\label{kmu}
\end{equation}
with the   MacDonald function satisfying  the differential equation
\begin{equation}
\big[- \frac{d^{2} }{d \xi^{2}} + (m^2+k_\perp^2)\, e^{2  \xi}-\omega^2\big]  K_{i\omega/\kappa}(z(\xi))=0.
\label{wxi}
\end{equation} 
Here and in the following we assume  dimensionful quantities to be given in units of powers of  $\kappa$. 
\vskip .1cm
{\bf Partition functions and  momentum-frequency relations}
\vskip .1cm
For calculating the  thermodynamic quantities,  we follow the procedure in \cite{SUUG94} and  restrict the  system under consideration to a part of the Rindler space.  The resulting  discrete  spectrum  consists of  eigenvalues characterized by 3 integers $n,n_{2},n_3$ and the basic thermodynamic quantity,  the partition function, is given by,
\begin{equation}
\ln Z = -\sum_{n, n_2,n_3}^\infty\ln \big(1-e^{-\beta\, \omega(n,n_2,n_3)}\big)\,.
\label{dos}
\end{equation}
In the transverse directions   the system is restricted to a square with side-length $\mathcal L$. We impose periodic boundary conditions  and replace the sum over $n_2,n_3$ by an integral, 
\begin{eqnarray}
\ln Z =- \frac{{\mathcal A}}{2\pi} \sum_{n=1}^\infty \int_{0}^\infty k_\perp\,dk_\perp\ln \big(1-e^{-\beta \,\omega_n(k_\perp)}\big)\,,\quad {\mathcal A}={\mathcal L}^2\,.
\label{lnz}
\end{eqnarray}
The restriction of the $\xi$ variable has to account for  the infinite degeneracy of the spectrum 
\cite{LOY08}.
Related to this degeneracy is the well known fact that the Minkowski ground state is seen by a uniformly accelerated observer as a  system at finite temperature, the  (Unruh) temperature,\,cf.\,\cite{UNRU76},
\begin{equation}
T=\frac{1}{\beta}=\frac{1}{2\pi}\,.
\label{rite}
\end{equation}
Various possibilities exist to remove the degeneracy.  We also follow here the procedure in \cite{SUUG94}  and remove the degeneracy by requiring the space to be limited to the region $\xi \ge \xi_0$.  We impose   Dirichlet boundary conditions for  the eigenmodes at the  finite distance $e^{\xi_0}$ from the horizon. 
Due to the exponential  increase of the repulsive ``potential'' in the wave equation (\ref{wxi}),   a discrete spectrum with respect to the $\xi$ variable is obtained without erecting a second wall.
For a given value of $\omega$ and given the number $n$ of zeroes, the vanishing of the MacDonald function, cf.\,Eq.(\ref{kmu}), determines
the value of $k_\perp$, 
\begin{equation}
 K_{i\omega}(\mathcal{K}_n(\omega))=0\,,\quad \text{with}\quad  \mathcal{K}_n(\omega)=e^{\xi_0}\sqrt{k_\perp^2(n,\omega)+m^2}\,.
\label{bocu}
\end{equation}
For evaluation of the thermodynamic quantities   the level density (cf.\,Eq.\,(\ref{lnz}))
associated with the transverse motion 
has to be computed. 
We shall refer to the resulting  relation between    ${\mathcal K}_n$ and  $\omega$ as ``momentum-frequency (m-f) relation'' (cf.\,the corresponding well known  m-f relation (\ref{dispm1}) in Minkowski space).  In terms of these quantities the level density can be computed for any value of the parameters $\xi_0,\,\mathcal{A},\,m$,
\begin{equation}
\frac{\mathcal A}{2\pi}\sum_{n=1}^\infty k_\perp(n,\omega) dk_\perp(n,\omega) = 
\frac{{\mathcal A}}{2\pi\ell^2} \sum_{n=1}^\infty {\mathcal K}_n(\omega) \frac{d  {\mathcal K}_n(\omega)}{d\omega} \theta\big({\mathcal K}_n(\omega)-m_\ell\big)\,d\omega\,,
\label{omk}
\end{equation}
where $\ell=e^{\xi_0}$ denotes the distance  of the boundary to the horizon   and $m_\ell$ the  mass of the field in units of $1/\ell$. 
According to Eq.\,(\ref{lnz}) the  logarithm of the partition function reads
\begin{equation}
\ln Z = 
\frac{\mathcal{A}}{4\ell^2}\, \sum_{n=1}^\infty \zeta_n(\beta,\omega_n^0)\,, 
\label{stpart}
\end{equation}
where, after an integration by parts,  the  functions $\zeta_n$
are  given by,
\begin{equation}
\zeta_n(\beta,\omega_n^0)=\frac{1}{\pi}\Big( {\mathcal K}_n^2(\omega_n^0)\ln\big(1-e^{-\beta \omega_n^0})+\beta\int_{\omega_n^0}^\infty  d\omega\, \Phi_n(\omega,\beta)\Big)\,,\quad \Phi_n(\omega,\beta)= \frac{{\mathcal K}_n^2(\omega)}{e^{\beta\omega}-1}\,,
\label{zet2}
\end{equation}
with the lower limit of the $\omega$-integration (cf.\,Eq.\,(\ref{bocu})),
\begin{equation}
{\mathcal K}_n(\omega_n^0)=m_\ell\,.
\label{o0n}
\end{equation}
In the following we assume the mass to vanish.   It will be shown in the last paragraph of Section III   that only under extreme  conditions finite mass effects can become relevant.
\vskip .1cm
{\bf m-f relations in Rindler and Minkowski space}
\vskip .1cm
The   m-f relations and partition function in Rindler space will be compared in  the following  with  the corresponding quantities  in Minkowski space.  To this end,  we assume a scalar massless field to be  confined in one direction  to an interval of size $\lambda$,\,i.e.,\,the Minkowski space  m-f relations
are given by, (for comparison, cf.\,Eq.\,(\ref{bocu})),
\begin{equation}
\mathcal{K}_{M, n}(\omega)= \sqrt{\omega^2-\big(n\pi/\lambda\big)^2}\,,
\label{dispm1}
\end{equation}
and the partition function reads (cf.\,Eq.\,(\ref{lnz}) and for comparison  Eqs.\,(\ref{stpart},\,\ref{zet2}))
\begin{eqnarray}
&&\hspace{4cm}\ln Z_M = \frac{\mathcal{A}}{4\lambda^2}\sum_{n=1}^\infty \zeta_{M,n}(\beta,\lambda), \nonumber\\
&&\quad \text{with}\quad \zeta_{M,n}(\beta,\lambda)=\frac{\beta\lambda^2}{\pi} \int_{n\pi/\lambda}^\infty\hskip-.3cm\,d\omega\,\Phi_{M,n}(\omega,\beta),
\;\;\Phi_{M,n}(\omega,\beta)=\frac{\mathcal{K}_{M,n}^2(\omega)}{e^{\beta \omega}-1}\,.
\label{partfM}
\end{eqnarray}
 The core of our numerical studies of the m-f relations are displayed in Fig.\,\ref{disprel}.  In a log-log plot are shown the  m-f relations in Rindler (${\mathcal K}_n(\omega)$) and in Minkowski space  (${\mathcal K}_{M,n}(\omega)$)  for various values of $n$. While the Minkowski space m-f relations are  given analytically by Eq.\,(\ref{dispm1}) the Rindler space m-f relations  have been   obtained by solving Eq.\,(\ref{bocu})  numerically.  
Also shown is the square root of the  ``Boltzmann factor'' which, if multiplied  with ${\mathcal K}_n(\omega)$, yields the  square root of the  integrands of the partition function, cf.\,Eq.\,(\ref{zet2}). The peculiarities of the Rindler space m-f relations are evident  in the comparison with the Minkowski m-f relations. In Minkowski space   each of the momenta $\mathcal{K}_{M,n}$ exhibits  a threshold which  for vanishing mass is given by  $\omega= n\pi/\lambda$  (cf.\,Eq.(\ref{dispm1})) 
while in Rindler space the m-f relations cover the whole range of $\omega$. This is a consequence of imposing in Rindler space only one boundary condition  (cf.\,Eq.(\ref{bocu})). Unlike in Minkowski space, a second boundary condition  is  not necessary due to the infinitely increasing strength of the repulsion (cf.\,Eq.(\ref{wxi})) with increasing $\xi$. In other words, the waves ``tunnel'' into a region which for instance would not be accessible to a classical particle.

\begin{figure}[h]
\begin{center}
\includegraphics[width=.6\linewidth]{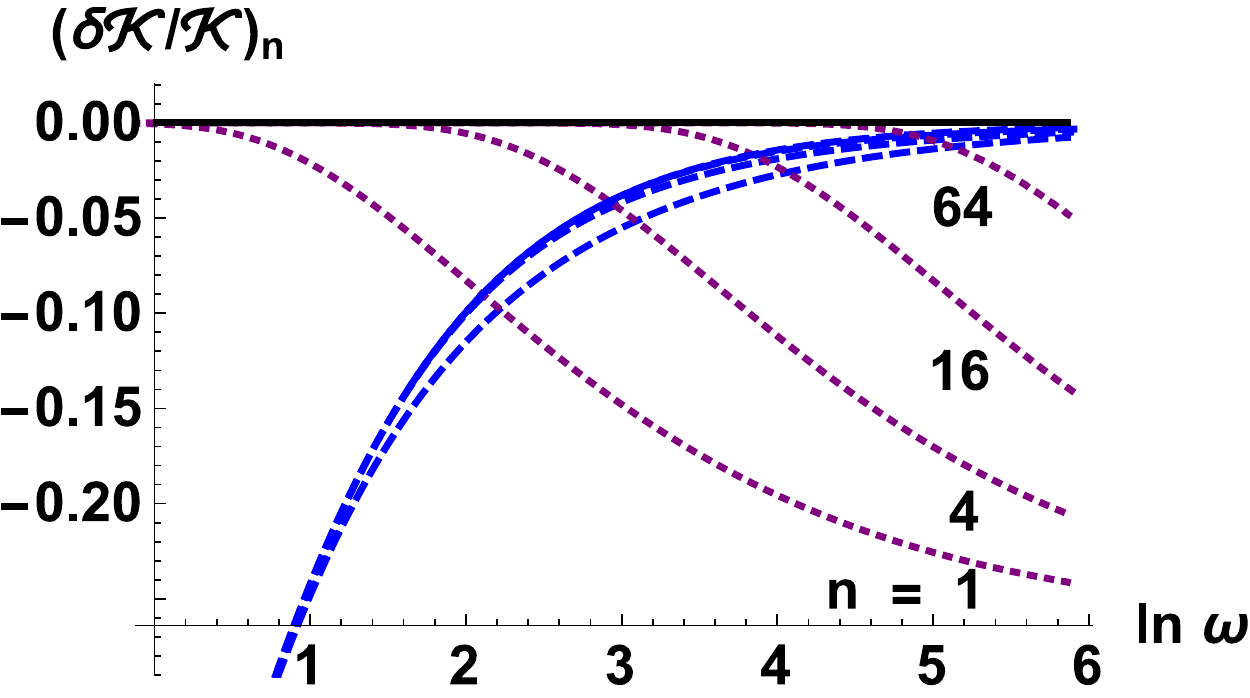}
\caption{Deviations  $\tilde{\mathcal K}_n/{\mathcal K}_n-1$ of the pole dominance approximation,    decreasing with increasing $\ln\,\omega$, (dotted, purple lines) and of the WKB approximations  $\hat{{\mathcal K}}_n/{\mathcal K}_n-1$, increasing with increasing  $\ln\,\omega$, (dashed, blue lines)
from the exact results. Asymptotically, the  deviations $\tilde{{\mathcal K}}_n(\omega)/{\mathcal K}_n(\omega)-1$ approach  the $\ln \omega$-axis $\delta\mathcal{K}/\mathcal{K}=2/e-1$, cf.\,Eq.\,(\ref{ltil}) and   the  deviations $\hat{{\mathcal K}}_n(\omega)/{\mathcal K}_n(\omega)-1$ approach 0, cf.\,Fig.\,\ref{disprel} and Eq.\,(\ref{lawkb}). For fixed $\omega$ the absolute values of the deviations (of pole dominance and WKB approximations)    decrease with increasing $n$.} 
\vskip -.5cm
\label{apdisprel}
\end{center}
\end{figure}

For assessing  the accuracy of the   numerically determined  zeroes of the MacDonald functions,
cf.\,Eq.\,(\ref{bocu}), a well defined  measure      is the  following  quantity, cf.\cite{RG65}, 
$$\chi(\omega,n)=\frac{K_{i\omega}(z)}{z\frac{d}{dz}K_{i\omega}(z)}=\frac{K_{i\omega-1}(z)-K_{i\omega+1}(z)}{i\omega(K_{i\omega-1}(z)+K_{i\omega+1}(z))}\,.$$
For $ -2.4\le \ln \omega\le 6.0$ (cf.\,Fig.\,\ref{disprel}), and for $n=1$ (with similar results for $n>1$) the accuracy varies in the interval,
$$
-42 \le \ln \chi(\omega,1) \le 
-32\,.$$  
Having determined the m-f relations one can proceed directly to Section III and  determine by integrations  the thermodynamic quantities. Before proceeding however,  we  will  develop   two different but  complementary  analytical approximations,  the ``pole dominance'' (PD) and the WKB approximation   in order to gain insight into the properties of the m-f relations.   
\vskip .1cm
{\bf Pole dominance approximation}
\vskip .1cm
In the small ${\mathcal K}_n/\omega$ regime, the m-f relations  are determined by the asymptotics, $\xi_0\ll-1,$ of  the  MacDonald functions, cf.\,\cite{RG65}, 
\begin{eqnarray*}
k_{i \omega} ({\mathcal K}_n) \approx  -\sqrt{\frac{2}{\pi}} \; \sin \big(\omega \xi_0 - \delta \big)\, , \quad
e^{2 i \delta} = \frac{\Gamma(1 + i\omega)}{\Gamma (1 - i \omega)}
\left ( \frac{e^{-\xi_0} {\mathcal K}_n(\omega)}{2 } \right ) ^{- 2 i \omega} \,.
\end{eqnarray*}
Thus, $k_{i \omega} ({\mathcal K}_n)$ vanishes if
$\omega\xi_0-\delta+n\pi=0,$
and we obtain,
\begin{equation} 
\ln\,{\mathcal K}_n(\omega)\approx  (\arg \Gamma(1+i\omega)-n\pi)/\omega+\ln 2\,\equiv\ln\,\tilde{{\mathcal K}}_n(\omega).
\label{kappanl}
\end{equation} 
The positions of the singularities of ${\mathcal K}_n$ in the complex $\omega$-plane coincide with  the positions of the poles of the Rindler space propagator Fourier transformed in time \cite{LOY11}, and we shall refer to  (\ref{kappanl}) as  pole dominance (``PD'') approximation. The accuracy of this approximation together with the WKB approximation as functions of the logarithm of $\omega$ for various values of $n$  is shown in Fig.\,\ref{apdisprel}. 

To analyze the shape of the m-f relations we first consider the small $\omega$ region 
where the m-f relation (\ref{kappanl}) simplifies, 
\begin{equation}
\tilde{{\mathcal K}}_n\underset{\omega \to 0}{\sim} e^{-(n\pi/\omega +\gamma-\ln 2)}\,,
\label{logom}
\end{equation}
with Euler's constant $\gamma$.  It exhibits an essential singularity  at $\omega=0$ which is responsible for the steep increase of  $\tilde{{\mathcal K}}_n$  with $\omega$. 
This  behavior   has its origin in the infinite degeneracy of the spectrum of the Rindler space Hamiltonian in the absence of the boundary (cf.\,the wave equation\,(\ref{wxi})) corresponding to a vertical line for each value of $\omega$. Furthermore, as in the degenerate case, in the regime of validity of Eq.\,(\ref{logom}),  the curves for different $n$ are parallel,\,i.e.,\,for $\tilde{{\mathcal K}}_m(\omega_m)=\tilde{{\mathcal K}}_n(\omega_n)$ these curves satisfy $\ln \omega_m -\ln\omega_n= \ln m-\ln n $\,.  The vertical distances between the curves (cf.\,Fig.\,\ref{disprel}) are given by Eq.\,(\ref{kappanl})\,,
\begin{equation}
\ln \tilde{{\mathcal K}}_m(\omega)-\ln \tilde{{\mathcal K}}_n(\omega)= (n-m)\pi/\omega\,,
\label{ratphi}
\end{equation}
in agreement within 10\%  with the numerical results for $\omega \le 8$ (cf.\,Fig.\,\ref{apdisprel}).  
 According to  Eq.\,(\ref{kappanl}),  the     m-f relations, evaluated in PD approximation, converge to the limit,
\begin{equation}
\tilde{\mathcal K}_n(\omega)\underset{\omega \to \infty}{\sim}\frac{2\omega}{e} e^{-(n-1/4)\pi/\omega},
\label{ltil}
\end{equation} 
which deviates  from the  numerically determined asymptotics  by the factor $2/e$. 
\vskip .1cm
{\bf WKB approximation}
\vskip .1cm
For not too small values of $\omega$\,(cf.\,Fig.\,\ref{apdisprel}), the WKB approximation\, \cite{tHOO85}, can be applied successfully for calculating the   m-f relations. Together with an additional approximation,  it has become the most common tool  for calculating analytically  thermodynamic quantities   in Schwarzschild \cite{tHOO85} and Rindler \cite{SUUG94} spaces (cf.\,also the reviews \cite{FRFU98} and \cite{PADM05}). The WKB m-f relations,\,$\hat{{\mathcal K}}_n(\omega)$,\, are obtained by solving the equation, cf.\,\cite{SUUG94},\,\cite{tHOO85},
\begin{equation}
n\pi=\int_{\xi_0}^{\xi_1} d\xi \sqrt{\omega^2-e^{2(\xi-\xi_0)}\hat{{\mathcal K}}^2_n(\omega)}\,,
\label{snp}
\end{equation}
where $\xi_1$ denotes the turning point. This integral can be evaluated analytically and the WKB m-f relations are obtained as solutions of the following equation,
\begin{equation}
\frac{n \pi}{\omega} = h(\eta)\,,\quad h(\eta)=- \sqrt{1-\eta^2}+  \ln \big(\sqrt{1-\eta^2}+1\big)-\ln \eta\,,\quad 0\le \eta=\frac{\hat{{\mathcal K}}_n}{\omega} \le 1\,.
\label{feta}
\end{equation} 
This equation implies 
that the solutions $\hat{{\mathcal K}}_n$  are given in terms of the inverse of the function, 
\begin{equation}
\hat{{\mathcal K}}_n(\omega)= \omega\, h^{-1}\Big(\frac{n\pi}{\omega}\Big)\,.
\label{rel2}
\end{equation}
Although a complete analytical solution cannot be attained, 
expansion of  $h(\eta)$ around $\eta=0$ and  $\eta=1$ yield explicit expressions  in the limits of small and of large 
values of  $\omega$ respectively, 
\begin{equation}
\hat{{\mathcal K}}_n(\omega) \underset{\omega \to 0}{\sim}\omega \,e^{-(n\pi /\omega +1-\ln 2)}\,,\quad \quad
\hat{{\mathcal K}}_n(\omega)\underset{\omega \to \infty}{\sim}\omega\Big(1+\frac{1}{2}\Big(\frac{3n\pi }{\omega}\Big)^{2/3}+\frac{7}{40}\Big(\frac{3n\pi }{\omega}\Big)^{4/3}\Big)^{-1}\,.
\label{lawkb}
\end{equation} 
As one might expect, the semiclassical  WKB approximation  fails to describe properly  the m-f relations in the small $\omega$ limit.  
Compared to the PD approximation $\tilde{K}_n(\omega)$ (\ref{logom}) which becomes exact in the small $\omega$ limit, the corresponding WKB approximation is suppressed (cf.\,Fig.\ref{apdisprel}) by the factor $\omega\,e^{-1+\gamma}$.      It reproduces  however, in agreement with  the numerical results,
the $\omega\to\infty$ limit of the m-f relation.

We observe that  for large values of $\omega$  the Rindler and Minkowski  space m-f relations still exhibit significant differences. While in Minkowski space, the momenta with different $n$ converge, they diverge in Rindler space (cf.\,Eq.\,(\ref{lawkb})) 
\begin{equation*}
{\mathcal K}_{M,n}(\omega)-{\mathcal K}_{M,m}(\omega)  \enskip\underset{\omega \to \infty}{\longrightarrow}\enskip \frac{1}{2\omega}\frac{\pi^2}{\lambda^2}(m^2-n^2),\quad \hat{{\mathcal K}}_n(\omega)-\hat{{\mathcal K}}_m(\omega)  \enskip\underset{\omega \to \infty}{\longrightarrow}\enskip
\frac{(3\pi)}{2}^{2/3}(m^{2/3}-n^{2/3})\,\omega^{1/3} \,.
\label{laom}
\end{equation*}
Fig.\,\ref{disprel} demonstrates this difference between Rindler and Minkowski space at  $\ln \omega \ge 4.6$. 
Related to this property is the difference of the level density at large $\omega$ (cf.\,Eq.(\ref{dispm1})),
\begin{equation}
\frac{1}{2}\frac{d}{d\omega} \hat{{\mathcal K}}_{n}^2(\omega) \enskip\underset{\omega \to \infty}{\longrightarrow} \omega - \frac{2}{3}(3n\pi)^{2/3} \omega^{1/3},\qquad \,
\frac{1}{2}\frac{d}{d\omega} {\mathcal K}_{M,n}^2(\omega)=\omega\,.
\label{KMK}
\end{equation}
\section{Thermodynamic quantities}
\begin{figure}[h]
\begin{center}
\includegraphics[width=.55\linewidth]
{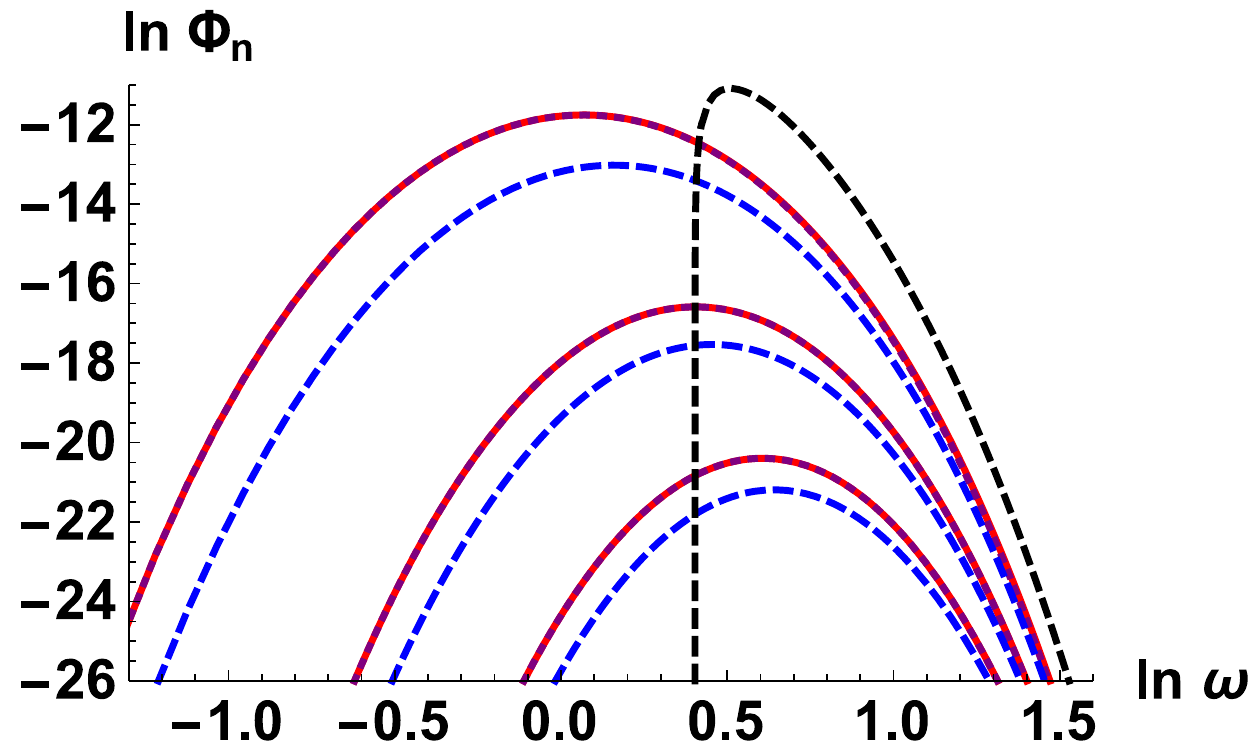}
\caption{Logarithm of the integrand $\Phi_n$ of the partition function (Eq.\,(\ref{zet2}))  as a function of $\ln \omega$ for  $n=1,2,3$\,,  
calculated  numerically (solid, (red) lines), in the  PD approximation (dotted, (purple) lines) and in the WKB approximation (dashed, (blue) lines)  in comparison  with  the   Minkowski-space quantity $\ln \Phi_{M,1}(\omega)$ (cf.\,Eq.\,(\ref{partfM})).}
\label{lnphi}
\end{center}
\vskip-.4cm
\end{figure}

{\bf Thermodynamic properties of massless  fields}
\vskip .1cm
Given the m-f relations, the thermodynamic quantities are, apart from a boundary term, (present only if $m\neq 0$)   obtained by integrating  ${\mathcal K}_n^2(\omega)$ over $\omega$ weighted with the ``Boltzmann factor'' (Eq.\,(\ref{zet2})) and by summing over $n$.  The integrands are shown in  Fig.\,\ref{lnphi} for $n=1,2,3$. These curves are the result of  the interplay between the increasing  squared momenta and the decreasing ``Boltzmann factor'' as a function of $\omega$ cf.\,Fig.\ref{disprel}. The positions of the maxima coincide roughly with the positions of the crossing of the squared momenta and of the ``Boltzmann factor'' in Fig.\ref{disprel}. These results  demonstrate  dominance of the $n=1$ contribution which in turn is  dominated by the maximum at $\omega\approx1$ with ${\mathcal K}_1\approx 0.06$ (cf.\,Fig.\,\ref{disprel}). Also shown is the integrand of the corresponding Minkowski space quantity  $\Phi_{M,1}(\omega)$ (cf.\,Eq.\,(\ref{partfM})) which,  due to the presence of threshold at $\omega_{th}=\pi/\lambda=\pi/2.1$ (cf.\,Fig.\,(\ref{disprel})),  exhibits a rather  different structure. The maximum is reached at   $\omega \approx  \omega_{th}+1/\beta$ and approaches  $\Phi_1(\omega)$  asymptotically (cf.\,Eqs.\,(\ref{partfM}), (\ref{lawkb}))
$$\frac{\Phi_{M,1}(\omega)}{\Phi_1(\omega)}\enskip\underset{\omega \to \infty}{\longrightarrow}1+\big(3\pi/\omega\big)^{2/3}.$$

Given these results for $\Phi_n(\omega)$, it is straightforward to determine numerically the logarithm of the partition function (cf.\,Eqs.\,(\ref{stpart},\ref{zet2}))  and the entropy, 
\begin{eqnarray}
S=  \frac{{\mathcal A}}{4\ell^2}(1-\beta\,\partial_\beta)\sum_{n=1}^\infty
\zeta_n(\beta,\omega_n^0)\Big|_{\beta=2\pi},
\label{thqu}
\end{eqnarray}
and  to study analytically these quantities. 
In the PD approximation the $n$-sum can be carried  out with the help of  the relation (\ref{ratphi}), 
and the calculation of  the logarithm of the partition function is  reduced to  a quadrature (cf.\,Eq.\,(\ref{kappanl})),
\begin{eqnarray}
\label{appd}
&&\hspace{-1.4cm}\ln \tilde{Z}
=\frac{\beta\mathcal{A}}{\pi\ell^2}\, \int_0^\infty\, \frac{e^{2\arg\,\Gamma(1+i\omega)/\omega}\,d\omega}{(e^{\beta\omega}-1) (e^{2\pi/\omega}-1)} \,.
\end{eqnarray}
Correspondingly, in the WKB approximation, Eq.\,(\ref{rel2}) implies,
\begin{equation}
\hat{{\mathcal K}}_n(\omega) = n\,\hat{{\mathcal K}}_1(\omega/n) \,,
\label{nsuaw}
\end{equation}
and the partition function can  be expressed in terms of $\hat{{\mathcal K}}_1(\omega)$
\begin{equation}\ln \hat{Z} =
\frac{\mathcal{A}}{4\ell^2} \int_0^\infty d\omega\,\hat{\sigma}(\omega),\;\; \text{with} \;\; \hat{\sigma}(\omega)= 
  \frac{\beta}{\pi}\, \hat{{\mathcal K}}_1^2(\omega) \sum_{n=1}^\infty  
\frac{n^3}{e^{\beta n \omega}-1}\,\,. 
\label{apwkb}
\end{equation}

\begin{table}[h]
\begin{center} 
\begin{tabular}{|r|r|r|r|r|r|}  \hline 
$\;\;$&$\zeta_1\;\;\;\quad$&$\sum_n\zeta_n$\;\;\;\;& \;\;\;$S\cdot 4\ell^2/\mathcal{A}\quad $ \\ \hline \hline 
NUM &$1.27\cdot 10^{-5}\;$ &$1.30 \cdot 10^{-5}$ &$9.68\cdot 10^{-5}\;\;$\\ \hline
PD\;\; & $ 1.26\cdot 10^{-5}\;$&$ 1.28\cdot 10^{-5} \; $&$  9.64\cdot 10^{-5} \;\;$\\ \hline
WKB&$3.72\cdot 10^{-6}\;$&$ 3.77\cdot 10^{-6}$\;  &$  3.08\cdot 10^{-5} $ \;
\\ \hline\hline
\end{tabular}
\end{center}
\caption{$n=1$  and summed contributions to the logarithm of the partition function and up to a factor the entropy $S$ (cf.\,Eq.\,(\ref{thqu})) for vanishing mass   calculated numerically  in the PD and WKB  approximations.}
\label{tdl1}
\end{table} 
The results of our studies of the thermodynamic quantities are compiled in Table \ref{tdl1}. Up to corrections of about 1\%, the $n\,=\,1$ terms of the logarithm of the partion function (and similarly of the entropy)    coincide with  the $n$-summed results reflecting the  strong suppression  of $\ln \Phi_n$ with increasing $n$ as displayed in Fig.\,\ref{lnphi}. The PD results agree with the corresponding  numerically determined ``exact'' results with an accuracy of better  than 1\%, cf.\,Fig.\,\ref{lnphi}, while the WKB results are too small  by a factor of about 3.
\vskip.3cm
{\bf Black holes as low temperature systems }
\vskip.3cm
Qualitative confirmation of and insights into the numerical results can be obtained by applying the approximate expression (\ref{logom}) according to which 
\begin{equation}
\Phi_n(\omega,\beta)\approx 4 e^{-2\gamma} e^{-2\pi n/\omega -\beta\omega}\,.
\label{phiap1}
\end{equation}
We conclude that the maxima of $\Phi_n(\omega,\beta)$  are given by 
\begin{equation}
\omega_n=\sqrt{2 \pi n/\beta},\quad n=1,2,\ldots ,
\label{phiap2}
\end{equation}
implying
\begin{equation*}
\ln \frac{\Phi_n(\omega_n,2\pi)}{\Phi_1(\omega_1,2\pi)}=-4\pi(\sqrt{n}-1) =-5.2,\;-9.2\quad \text{for}\quad n=2,3\,. 
\label{phiap3}
\end{equation*}
These values of the   positions and ratios of the maxima agree well with the numerical results shown in Fig.\,\ref{lnphi}. Furthermore, by carrying out the $\omega$-integration we obtain (cf.\,Eq.\,(\ref{zet2})),
\begin{equation*}
\frac{\zeta_n}{\zeta_1}\approx n^{1/4}\,e^{-4\pi(\sqrt{n}-1)}\approx 0.0065,\;0.00013 \quad \text{for} \quad n=2,3\,,
\label{phiap4}
\end{equation*}
which, within 10 and 25 \%, agree with the corresponding  numerical results. The partition function is dominated by the $n=1$ term with
\begin{equation*}
\zeta_1\approx 2^{5/2} e^{-2\gamma}  e^{-4\pi}=6.22\cdot 10^{-6}\,,
\label{phiap5}
\end{equation*}
which agrees within a factor of 2 with the numerical results in Table \ref{tdl1}. In the language of thermodynamics, the  dominance of the $n=1$ term is to be interpreted as the low temperature limit. Only by  a sufficient increase of the temperature $T\gg 1/2\pi$,\,i.e.,\,by decreasing the slope of the Boltzmann factor in Fig.\ref{disprel}, the modes with $n>1$ can contribute significantly. According to  Eq.\,(\ref{phiap2}), this is the case if   the temperature  increases  by a factor of  $(4\pi(\sqrt{n}-1))^2 $.  

This dominance of the $n=1$ contribution is not at all specific for the Rindler space or the other spaces to be considered.  
For sufficiently large values of $\upsilon=\beta\pi/\lambda$  which is the case for the choice $\lambda=2.1$, (cf.\,Fig.\,\ref{disprel}),  the Minkowski space partition function is given by (cf.\,Eq.\,(\ref{partfM})),
\begin{equation}
\ln Z_M\approx\frac{\mathcal{A}}{\lambda^2}\sum_{n=1}^{\infty}
\frac{\pi(1+n\upsilon)}{2\upsilon^2}e^{-n \upsilon}\,,
\label{Mik}
\end{equation}
i.e.,\,the sudden increase of $\Phi_{M,n}(\omega)$ is due to the thresholds at $\nu_n=n\pi/\lambda$    
and is dominated by 
$\zeta_{M,1}$. 
\vskip .3cm
{\bf Modified WKB approximation and analytical expressions for partition function and entropy}
\vskip .1cm
Since the work of 't Hooft and of Susskind and Uglum  a simplified version of the WKB approximation has been the commonly used tool for calculating analytically the black hole thermodynamic quantities. This simplification is valid only if two conditions are satisfied: The sum over the modes in Eq.\,(\ref{apwkb}) can be replaced by an integral and the range of this integration can be increased  by changing the lower limit of the integration  from 1 to 0,\,i.e.\,,
\begin{equation}
\ln \hat{Z} \to \ln \hat{Z}_{apx} =  
\frac{\beta\mathcal{A}}{4\pi\ell^2} \int_{0}^\infty d\nu\,\nu^3 \hspace{-.15cm}\int_0^\infty\hspace{-.1cm} d\omega \frac{1}{e^{\beta\nu \omega}-1}\,\hat{{\mathcal K}}_1^2(\omega)\,.
\label{cond}
\end{equation}
After carrying out the $\nu$ integration, also the $\omega$ integral can be calculated analytically with the help of Eqs.\,(\ref{feta}) and (\ref{rel2}),
\begin{equation}
 \int_{0}^\infty \frac{\nu^3}{e^{\nu \beta  \omega}-1}\,d\nu=\frac{1}{240}\Big(\frac{2\pi}{\beta\omega}\Big)^4\,,\hspace{1cm}\int_0^\infty\,d\omega \frac{\hat{{\mathcal K}}^2_1(\omega)}{\omega^4}=-\frac{1}{\pi}\int_0^1 d\eta\, \eta^2\, \frac{dh}{d\eta}=\frac{1}{3\pi}\,,
\label{nuint}
\end{equation}  
and yields  the following well known expression for the logarithm of  partition function  and  entropy,
\begin{equation}
\ln \hat{Z}_{apx}=\frac{{\mathcal A}\,/4\ell^2}{360\,\pi}\equiv\frac{1}{4}\hat{S}_{apx}\,, 
\label{wkbwr}
\end{equation} 
in agreement with  the results in \cite{tHOO85},\,\cite{SUUG94} (provided  $\ell$ is identified with the Planck length). However  they are   in   disagreement with the WKB results in  Table \ref{tdl1} which are smaller by a factor of 235 and 116 for partition function and entropy  respectively. 
The increase of the partition function by  two orders of magnitude   if the range of $\nu$ is extended from 1 to 0 does not come as a surprise. An increase of similar strength occurs in the increase of $\Phi_n$ when decreasing $n$ from 2 to 1 (cf.\,Fig.\ref{lnphi}) which in turn  has it's origin in the increase of both, the square of the momentum  $\mathcal{K}_\nu(\omega)$ and  the ``Boltzmann factor'' $1/(e^{\beta\omega}-1)$\,(cf.\,Fig.\ref{disprel}).  As shown in the Appendix, the value of $\ln\hat{Z}_{apx}$ (Eq.\,(\ref{wkbwr})) is obtained within $0.1\%$ if  the $\nu$ integration, (cf.\,Eq.\,(\ref{cond})) is limited to the  unphysical region $0\le \nu\le 1$.

The same phenomenon occurs if we proceed to calculate the Minkowski space partition function (\ref{Mik})  in a similar way and approximate the $n$-sum by an integration with the result, 
\begin{equation}
\ln Z_{M,apx}=\frac{{\mathcal A}}{\lambda^2}\int_{\nu_0}^\infty\hspace{-.3cm} d\nu\,\zeta_\nu^\mu =\frac{{\mathcal A}}{\lambda^2}\frac{\pi(2/\upsilon+\nu_0)}{2\upsilon^2}e^{-\upsilon \nu_0}\,.
\label{Mkin}
\end{equation}
With the values  $\nu_0=0$ and $\nu_0=1$, the ratio of the exact (Eq.\,(\ref{Mik})) and the approximate results are given, for $\upsilon \gg 1$, by
\begin{equation}
\frac{\ln Z_{M,apx}}{\ln Z_M}\Big|_{\nu_0=0} \approx \frac{2}{\upsilon^2}e^\upsilon\,,\quad \frac{\ln Z_{M,apx}}{\ln Z_M}\Big|_{\nu_0=1} \approx \frac{1}{\upsilon}\,, 
\label{rat}
\end{equation}
i.e.,\,we find the same pattern as above.   Also the  Minkowski space partition function is overestimated by orders of magnitude if  $\nu_0=0$ and much closer to the exact results if $\nu_0=1$. 

Qualitatively, the failure of replacing the sum over  either the  Rindler or the Minkowski space modes, by an integration is evident in view of Fig.\,\ref{disprel}. Only if, as a function of $\omega$, the ``Boltzmann factor'' is significantly flatter the replacement of the summation by an integration can be justified. In turn, this can be achieved  only by increasing the temperature,\,i.e.,\,by decreasing $\beta$ significantly which however is not an option  for the  black hole thermodynamics where $\beta=2\pi$. 
\vskip .1cm
{\bf The role of the boundary condition.}
\vskip .1cm
In concluding our discussion of the thermodynamic quantities of massless fields,   we discuss the role of the boundary condition for the thermodynamic quantities. 
We have seen  that the only way to vary the thermodynamic quantities is via the  prefactor  $\mathcal{A}/4\ell^2$. As will be shown in section IV, the expressions for the thermodynamic quantities   apply also for   de Sitter space and more generally for spherically symmetric spaces with static metrics. The only freedom which is left is the choice of the boundary condition. While  irrelevant, cf.\,\cite{FRFU98},  if many modes contribute,
the choice of the boundary condition  becomes important  if only one or a few  modes dominate. We demonstrate  this uncertainty  by modifying the brick wall model and replace the Dirichlet (Eq.\,(\ref{bocu})) by the Neumann boundary condition which in PD amounts to replace in Eq.\,(\ref{kappanl}) $n\pi/\omega \to (n-1/2)\pi/\omega$  resulting in the following expression for the partition function,
\begin{equation}
\ln \tilde{Z}_{Ne}
=\frac{\beta\mathcal{A}}{\pi\ell^2}\, \int_0^\infty\, \frac{e^{(2\arg\,\Gamma(1+i\omega)+\pi)/\omega}\,d\omega}{(e^{\beta\omega}-1) (e^{2\pi/\omega}-1)}\,. 
\label{dineu}
\end{equation}
Numerical evaluation of this expression yields, in comparison with the results (PD) of Table \ref{tdl1}  an enhancement of the partition function by a factor of 32 and of  the entropy  by a factor of 22.  Taking  into account that the numerical and the PD values of partition function and  entropy are larger than the WKB results (cf. Table \ref{tdl1}) by factors of 3.4 and 3.1 respectively, we find that up to factors of 2.1 and 1.7  the values of Eq.\,(\ref{wkbwr}) are obtained by imposing Neumann boundary conditions,\,cf.\,Eq.\,(\ref{dineu}).   
\vskip .1cm
{\bf Effects of masses on thermodynamic properties}
\vskip .1cm
With introduction of a  mass, a new scale enters, (for a calculation of the free energy for massive fields within the WKB approximation   cf.\,\cite{KAST94}). The distance $\ell=e^{\xi_0}$ to the horizon  appears not only as   prefactor $\ell^{-2}$  in the partition function. According  to Eq.\,(\ref{zet2})  the thermodynamic properties of both massive and massless fields are determined by  the  same quantities, ${\mathcal K}_n(\omega)$. Differences result exclusively from the surface terms, $\sim {\mathcal K}_n^2$, and the presence of the non-vanishing lower limit $\omega_n^0$ in  the $\omega$ integration which is determined by $m_{\ell}$,  
the product of the mass and the distance to the horizon, cf.\,Eq.\,(\ref{o0n}). As indicated  by Fig.\,\ref{lnphi}, the effect  of the non-vanishing surface term and of the lower limit of integration for $n=1$ is, at the level of 1\% or smaller,  negligible for $\omega \le 0.5$. Therefore  the minimal mass,   necessary for affecting the thermodynamic quantities,   must satisfy (cf.\,Eq.\,(\ref{o0n})),
\begin{equation*}
 m \ge {\mathcal K}_1(0.5)\,\,\ell^{-1}\, =  2\pi\, {\mathcal K}_1(0.5)\, T_{\ell}\,,
\label{plle0}
\end{equation*} 
where $T_{\ell}$ denotes the Tolman temperature at the distance $\ell$ from the horizon.  Identifying $\ell$ with the Planck length the above inequality  reads
in terms of the Planck mass $M_P$
\begin{equation}
m\ge 2\cdot 10^{-3} M_P\,.
\label{plle}
\end{equation}
Thus unless there are particles with masses of the order of at least $10^{-3}\,\times $ the Planck mass (``Micro black holes''),  or there is  a reason to increase the distance of the boundary to the horizon from the  Planck length to  at least  $10^{-3}\times $ Compton wavelength of the corresponding particle, the mass of the particle does not affect the thermodynamic quantities.

Given  the  independence of the thermodynamic quantities from the mass of the particles outside the range (\ref{plle}) we can get a rough estimate 
 of the  entropy generated by the particles of the standard model. To this end,  we assume that, apart from the multiplicity, all fundamental  particles (leptons, quarks, gauge bosons and the Higgs particle)   contribute the same amount to the entropy resulting in a value of $s(0)$ (cf.\,Table \ref{tdl1})  of the order of $10^{-2}$. The ambiguity  in the choice of the boundary condition results in  an uncertainty of a factor of 22 (cf.\,Eq.\,(\ref{dineu})). Together with the uncertainty in choosing  the value of the distance to the horizon,  this  estimate  could be wrong by one or two orders of magnitude. Beyond this uncertainty  we also have to take into account that  all     particles beyond the standard model with masses, in the range,
$$1\,\text{TeV}\le m\le 10^{13}\, \text{TeV}\,,$$ 
for instance, possible   supersymmetric partners of the particles in the standard model,  contribute with the same weight to the partition function and other thermodynamic quantities as  the particles with masses below 1\,TeV.
\section{Angular momentum - frequency  relations in de Sitter space}\label{dprrdS}
The purpose of the following study is twofold. We will first determine the angular momentum-frequency relations (am-f relations) in de Sitter space by applying analytical and numerical methods and at the same time we will establish quantitatively the connection between the Rindler space  m-f  and de Sitter space am-f relations. Thereby we will exhibit  quantitatively validity  and limits of the near horizon approximation.
\vskip .1cm
{\bf  am-f relations in de Sitter space}
\vskip.1cm
Starting point of our studies is the de Sitter space metric in static coordinates,\,cf.\,\cite{SPSV11}
\begin{equation}
ds^2=(1-r^2\kappa^2)dt^2-\frac{1}{1-r^2\kappa^2}dr^2-r^2d\Omega^2,
\label{desimet}
\end{equation}
with the   de Sitter radius given by the inverse of the surface gravity $\kappa$.
The radial eigenfunctions $\varphi_l(r)$ with angular momentum $l$ of the wave equation associated with the above metric   are well known \cite{LOPA78}.
We impose
the same type of boundary condition   as for the  Rindler space eigenfunctions,
\begin{eqnarray}
&&
\hspace{-.8cm}\varphi_l(r)\Big|_{r^2=1-e^{2\xi_0}}=0\,,\nonumber\\
&&
\hspace{-.8cm}\varphi_l(r)=r^{l}(1-r^2)^{\frac{1}{2}i\omega}\,_2F_1\,\big(({\mathcal K}_+ e^{-\xi_{0}}+i\omega)/2,\,({\mathcal K}_- e^{-\xi_{0}}+i\omega)/2;\,l+3/2;r^2\big) \,,
\label{hypf2}
\end{eqnarray}
where we have introduced, 
\begin{equation*}
{\mathcal K}_{\pm}=e^{\xi_{0}}\,\big(l+3/2\pm \sqrt{\,9/4-m^2}\,\big) \,.
\label{kapm}
\end{equation*}
In accordance with Eq.\,(\ref{bocu}), we have included  in this definition the suppression factor $e^{\xi_0}$ accounting for the time dilation of the transverse motion which, close to the horizon,  affects the  de Sitter space am-f  and the Rindler space m-f relations in the same way. We therefore  expect  the relevant  values of $l$ to  be of the order of $e^{-\xi_0}$.

By applying   one of the  linear transformation formulas for the hypergeometric function \cite{ABST65} the  change in Eq.\,(\ref{hypf2}) from $1-e^{2\xi_0}$  to the more appropriate  variable $e^{2\xi_0}$ is achieved, 
\begin{eqnarray}
\varphi_l\big(\sqrt{1-e^{2\xi_0}}\big)=\big(1-e^{2\xi_0}\big)^{l/2}\,\text{Re}\big[\rho(\omega,{\mathcal K}_+,{\mathcal K}_-,\xi_0)\, \sigma(\omega,{\mathcal K}_+,{\mathcal K}_-,\xi_0)\big]\,,
\end{eqnarray}
with
\begin{eqnarray}
&&\rho(\omega,{\mathcal K}_+,{\mathcal K}_-,\xi_0)=
\frac{\Gamma\big(({\mathcal K}_+ +{\mathcal K}_-) e^{-\xi_0}/2\big)\,\Gamma\big(-i\omega\big)\,e^{i\omega\xi_0}}{\Gamma\big(({\mathcal K}_+ e^{-\xi_0}-i\omega)/2\big)\Gamma\big(({\mathcal K}_- e^{-\xi_0}-i\omega)/2\big)}\,,\nonumber\\
&&\sigma(\omega,{\mathcal K}_+,{\mathcal K}_-,\xi_0) =\,_2F_1\Big(({\mathcal K}_+ e^{-\xi_0}+i\omega)/2,\big({\mathcal K}_- e^{-\xi_0}+i\omega)/2;i\omega+1;e^{2\xi_0}\Big)\,,
\label{thva}
\end{eqnarray}
and the boundary condition is rewritten as 
\begin{equation}
\psi\big(\omega,{\mathcal K}_+,{\mathcal K}_-,\xi_0\big)=\arg\,\rho\big(\omega,{\mathcal K}_+,{\mathcal K}_-,\xi_0\big)+\arg\,\sigma\big(\omega,{\mathcal K}_+,{\mathcal K}_-,\xi_0\big)=-\Big(n-\frac{1}{2}\Big)\pi\,.
\label{cdt}
\end{equation} 
Solution of this equation which we have carried out numerically   yields the de Sitter space am-f relations. 
\vskip .1cm
{\bf Near Horizon Approximation}
\vskip .1cm
For analytical studies,  this equation serves as  starting point for  the ``near horizon approximation'' which we define as the expansion in terms of the distance  $e^{\xi_0}$ to the horizon. 
To leading and next to leading    order we obtain from Eq.\,(\ref{thva}),  by treating $l$ as a continuous variable, 
\begin{equation}
\arg\,\rho_0
=\arg\Gamma(-i\omega) + \omega \ln ({\mathcal K}_{dS}/2)\,,\quad \arg\,\rho_1
=-\omega\,\frac{{\mathcal K}_++{\mathcal K}_-}{2{\mathcal K}_+{\mathcal K}_-}\, e^{\xi_{0}}=-\omega\,\frac{l+3/2}{l^2+3l+m^2}\,,
\label{lsrho}
\end{equation}
and after a tedious calculation, 
\begin{equation}
\arg\,\big(\sigma_0+\sigma_1\big) \approx \arg\,\sigma_0-\frac{{\mathcal K}_++{\mathcal K}_-}{2{\mathcal K}_{dS}}\, e^{\xi_{0}}\text{Im}\Big(  
I_{i\omega+1}({\mathcal K}_{dS})/I_{i\omega}({\mathcal K}_{dS})\Big)\,,
\label{sione}
\end{equation}
where
\begin{equation}
{\mathcal K}_{dS}=\sqrt{{\mathcal K}_+{\mathcal K}_-}=e^{\xi_0}\sqrt{l(l+3)+m^2}\,.
\label{kaka}
\end{equation} 
For $\omega\le 3$ the contribution of $\sigma$ to $\psi$ is negligible and the am-f relations satisfy,
$$ \arg\Gamma(-i\omega) + \omega \ln ({\mathcal K}_{dS}/2)=-\Big(n-\frac{1}{2}\Big)\pi\,.$$
These solutions coincide with the Rindler space am-f relations obtained in the PD approximation (\ref{kappanl}),\,i.e.,\,for given $n$ and $\omega$  the  identity, 
${\mathcal K}_{dS}(n,\omega)={\mathcal K}_n(\omega),$ 
holds.
As in Rindler space, the appearance of $\Gamma(-i\omega)$   reflects  the presence of poles (in the complex $\omega$ plane)  of the de Sitter space propagator Fourier transformed in time. 

\begin{figure}
\begin{center}
\includegraphics[width=.47\linewidth]{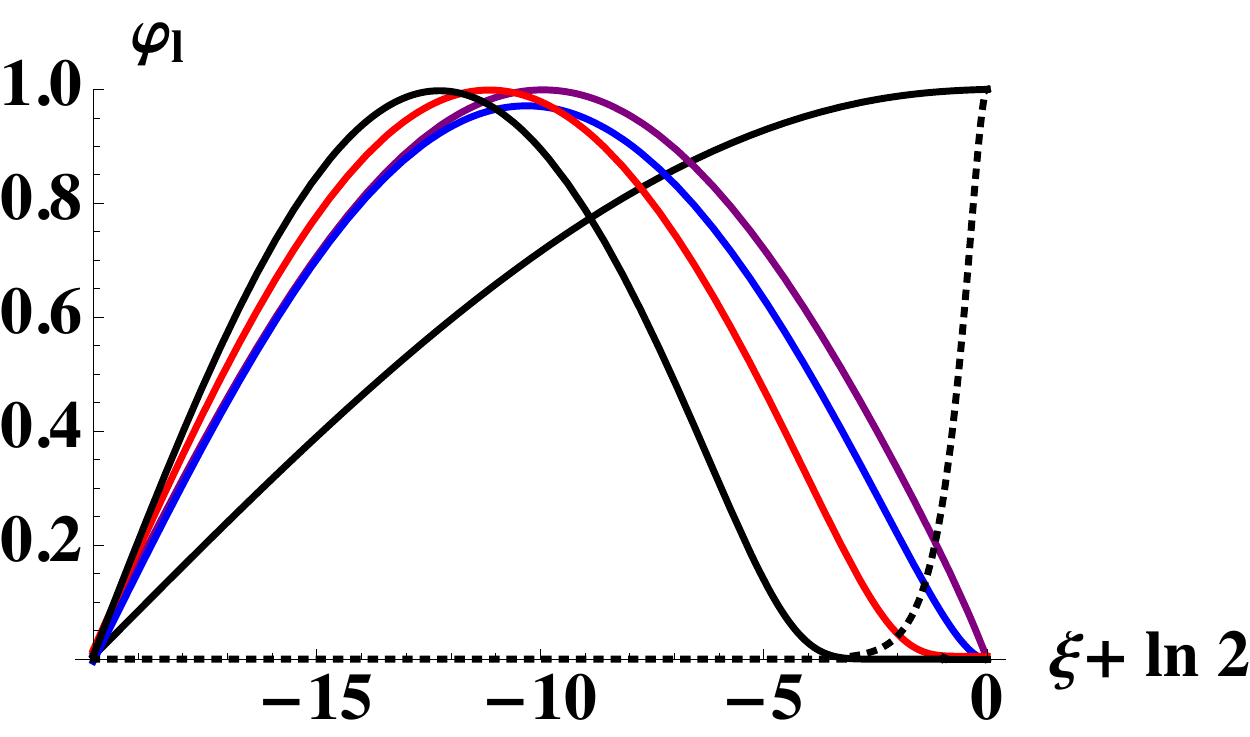}
\includegraphics[width=.5\linewidth]{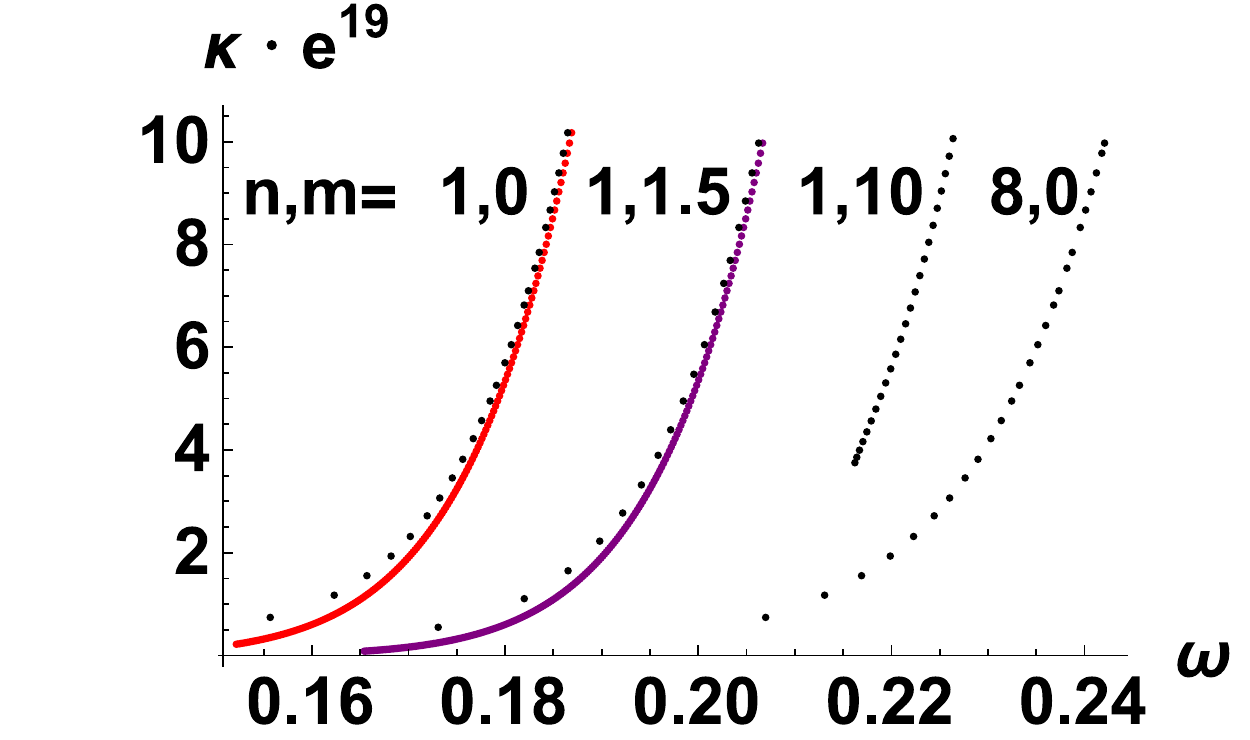}
\caption{Left: Absolute value of the de Sitter wave functions (\ref{hypf2})  as a function of $\xi$ for $m=0$,
and  $l=0,1,2,10,100$ with the maxima  normalized to 1. Dashed  curve:  $\delta V$,\, Eq.\,(\ref{dtvi}).  
Right: de Sitter space am-f relations (\ref{cdt}) for  $\xi_0=-20$, (relevant for the inflationary universe)  $l\le 26$  (black dots)  in comparison  with the Rindler space m-f relations (solid lines) with  ${\mathcal K}(\omega)$,\,Eq.\,(\ref{kaka}). 
}
\label{desitwffig}
\end{center} 
\vskip -.8cm
\end{figure}

To calculate  $\sigma_0$ we order the terms in the hypergeometric function (\ref{thva})  according to  powers of $e^{\xi_{0}}$  
and find to leading order  in this expansion  cf.\,\cite{RG65}, 
\begin{equation}
\hspace{.1cm}\sigma_0\big(\omega,{\mathcal K}_+,{\mathcal K}_-,\xi_0\big)\hspace{-.1cm}=\hspace{-.1cm}\sum_{n=0}^\infty \frac{1}{n!}\Big(\frac{{\mathcal K}_{dS} }{2}\Big)^{2n}\hspace{-.2cm} \frac{\Gamma(i\omega+1)}{\Gamma(i\omega+n+1)}=i\omega\Gamma(i\omega) ({\mathcal K}_{dS}/2)^{-i\omega}\,I_{i\omega}\big({\mathcal K}_{dS}\big).\hspace{-.57cm}
\label{apr1}
\end{equation}
Combining with the leading order term\,$\rho_0$  (Eq.\,(\ref{lsrho})), the approximate boundary condition  (cf.\,Eq.\,(\ref{cdt})) reads,
\begin{equation}
\psi_0\big(\omega,{\mathcal K}_+,{\mathcal K}_-,\xi_0\big)=
 \text{arg}\, I_{i\omega}\big({\mathcal K}_{dS}\big) +\frac{\pi}{2}= -n\pi+\frac{\pi}{2}\,.
\label{phi0}
\end{equation}
The relation \cite{RG65},
$$K_{i\omega}({\mathcal K})= \frac{\pi}{\sinh \omega \pi} \text{Im}\, I_{i\omega}({\mathcal K})\,,$$
implies that the zeroes of $K_{i\omega}({\mathcal K})$ coincide with the zeroes of $\text{Im}\,I_{i\omega}({\mathcal K})$. Thus to leading order, the am-f relations and therefore the thermodynamic quantities in de Sitter space coincide with 
the corresponding quantities in  Rindler space 
with the  identification of the ``momenta'' given in Eq.\,(\ref{kaka}).
\vskip .1cm
{\bf Validity and limitation of the  de Sitter - Rindler space connection}
\vskip .1cm
Origin as well as   limitations of the connection  between de Sitter space am-f and Rindler space m-f relations  are  easily identified  by comparing  the corresponding wave equations. To this end  we  change the de Sitter space coordinate $r=-\tanh(\xi+\ln 2)\,,\quad \xi\le -\ln\,2,$ 
which yields 
the following wave equation,
\begin{equation}
\Big[-\frac{d^2}{d\xi ^2} + V_l (\xi+\ln 2) - \omega^2\Big] \varphi_l(\xi) = 0,  \quad {\rm{with}} 
\quad V_l (\xi) = \frac{l(l+1)}{\sinh^2 \xi} +\frac{m^2-2}{\cosh^2 \xi}  
\label{pot}\,.
\end{equation}
Equation (\ref{pot}) can be replaced by  the Rindler space wave equation  (Eq.\,(\ref{wxi})) provided $\varphi_l(\xi)$ is localized sufficiently close  to the horizon. For this to happen the centrifugal barrier has to be sufficiently large,  
$\big(l(l+1)+m^2-2\big) \,e^{2\xi} \gg \omega^2\,.$  
The tighter and tighter   localization with increasing angular momentum is illustrated in  Fig.\,\ref{desitwffig}. Also shown  is the  relative difference between de Sitter and Rindler space ``potentials''
\begin{equation}
\delta V=(V_l(\xi+\ln 2)- 4\big(l(l+1) +m^2)e^{2\xi})/V_l(\xi+\ln 2)\,, 
\label{dtvi}
\end{equation}
for $l=5,\,m=0$.
Already for values with angular momenta as small as $l=1$,  only a weak overlap between the eigenfunction $\varphi_1$ and the difference $\delta V$  is found. On the other hand,  due to the absence of the centrifugal barrier  the wave-function for $m=0$ and $l=0$ is not dominated by the near horizon region. 

These considerations lead us  to consider in detail the am-f relation at small  angular momenta $l$ where significant differences between de Sitter and Rindler space results occur.   On the right hand side of Fig.\,\ref{desitwffig} are shown the discrete eigenvalues  for $n=1$  and $m=0,\,1.5,\,10$ (first 3 curves) and $n=8,\,m=0$ together with the corresponding  Rindler space m-f relations. The energies $\omega$ of the  second and third am-f  and m-f relations are shifted  by 0.02 and 0.04 respectively and ${\mathcal K}(\omega)$ is reduced by a factor of 6 for the  $n=8$ am-f and m-f relations. Not included in the figure are the eigenvalues for $l=0$ with exact values  $0.0798,\,0.119,\,0.176,\,1.18$ and approximate values $0,\,0.147,\,0.176,0\,$. As suggested by  the behavior of the wave functions, the de Sitter am-f relations approach fast the Rindler space m-f relations  with increasing $l$ and/or $m$. Already for $l=1,\,m=0$  Rindler and de Sitter space results (cf.\,Eq.\,(\ref{kaka}))  agree within 4\%. The discrepancy  can be reduced to 1\% by including the next to leading order (cf.\,Eqs.\,(\ref{lsrho}),\,(\ref{sione})) in  the near horizon approximation.
\vskip .1cm
{\bf Analytical expression for the l=0 contribution to the partition function}
\vskip .1cm
For $l=0,\,m=0$ (and $\omega\neq 0$),  the near horizon approximation fails.
The correspondence (\ref{kaka})  assigns to $l=m=0$  the Rindler space value   $\tilde{\mathcal K}=\omega=0$ (cf.\,Eq.\,(\ref{logom})) independent of the value of $n$ and the  next to leading order  (\ref{lsrho})  of the near horizon approximation diverges.  However,
 a closed expression for the  am-f relation can be obtained by applying  the duplication formula for the $\Gamma$-functions  in  (\ref{thva}), 
\begin{equation}
 \omega( \xi_0 -\ln 2) +(1-\delta_{l=0})\sum_{l^{\prime}=0}^{l-1}\arg(l^\prime+i\omega)+\arg(l+1+i\omega)=(-n+1/2)\,\pi\,. 
\label{dupfo}
\end{equation}  
For $l=0$ and not too large values of $n$, the approximate am-f relations are given by, 
\begin{equation}
\omega_n = -\frac{n-1/2}{\xi_0+1-\ln2}\,\pi\,,
\label{sm0o}
\end{equation}
which reproduces the exact values for $n\le 8$ with an accuracy of 1\% or better. 
On this basis also  the $l=0$ contribution to the partition function  can be calculated analytically   with the result
\begin{equation*}
\ln Z_{l=0} = -\sum_{n=1}^\infty\ln\bigg(1-\exp\Big(\frac{(2n-1)\pi^2}{\xi_0+1-\ln2}\Big)\bigg) = 1.32\,,
\end{equation*}
which is larger than $\sum_n\zeta_n$ (first line of Table\,\ref{tdl1}) by a factor of $4.1\cdot 10^5$. The relevance of  this contribution  depends on the density of states which  is $1$  for  $l=0$  and ${\mathcal A}/4\ell^2$ (cf.\,Eq.\,(\ref{stpart})) for large $l$.
\section{Angular momentum-energy relations in static spherically symmetric spaces close to horizons}\label{4shor}
The connection between the am-f and m-f relations  of scalar fields in  de Sitter and in Rindler space respectively can be generalized to a larger class  of static spherically symmetric spaces with a non-extremal horizon. Besides the de Sitter space,  the Schwarzschild,  the Schwarzschild/AdS or the Reissner-Nordstr\"om space  belong to this class.  The common structure of the   line element of this class of spaces reads cf.\,\cite{PADM05},
\begin{equation}
ds^2 = f(r) dt^2 - \frac{dr^2}{f(r)} - r^2d\Omega^2\,,
\label{liel}
\end{equation}
with the function $f(r)$ vanishing  at $r=r_0$ and, close to the horizon, is approximately given by, 
\begin{equation}
f(r)\approx (r-r_0)\,f^{\prime}(r_0)\,,\quad \text{with}\quad 
|(r-r_0) f^{\prime}(r_0)| = e^{2\kappa\xi}\,,\quad\text{and}\quad 2\kappa =\big|f^{\prime}(r_0)\big|\,.
\label{fr}
\end{equation}
The approximate metric (\ref{liel})  reads,
\begin{equation}
ds^2 \approx  e^{2\kappa\xi} (dt^2 - d\xi^2) - r_0 ^2\Big(1+\frac{1}{2\kappa r_0}e^{2\kappa\xi}\Big)^2d\Omega^{\,2}\approx e^{2\kappa\xi} (dt^2 - d\xi^2) - r_0 ^2\,d\Omega^{\,2}\,.
\label{spri}
\end{equation} 
The last step of the approximation is valid only if the radial eigenfunctions are concentrated in the region close to the horizon which is never the case for vanishing angular momentum $l$ and mass $m$. The approximate metric (\ref{spri}) 
is  the metric of a product space of the $1+1$ Rindler space and the 2-sphere. Comparison of this metric with the Rindler space metric (\ref{rin}) shows that (up to the normalization) the eigenfunctions are given by the MacDonald functions, which vanish at the boundary (cf.\,Eq.\,(\ref{bocu})),
\begin{equation}
\label{spmcd}
K_{i\omega}({\mathcal K}_{sp})=0\,,\; \;\; {\mathcal K}_{sp}=e^{\xi_0}\sqrt{m^2+l(l+1)/r^2_0}\,,
\end{equation}
with $\xi_0,\,r_0$ and $m$ given in units of $1/\kappa$ and $\kappa$ respectively. 
At this point we can proceed as above in identifying the m-f and am-f relations  of Rindler and spherical Rindler space respectively. 

To test the range of validity of this type of ``near horizon approximation'', we apply the  above approximation to de Sitter space where, according to Eq.\,(\ref{fr}), 
\begin{equation}
r_0=1/\kappa\,,
\label{DSM}
\end{equation}
and the near horizon metric and ${\mathcal K}_{sp}$ are given by,
\begin{equation}
ds^2 = e^{2\xi} (dt^2 - d\xi^2) - \,d\Omega^{\,2}\,,\quad 
 {\mathcal K}_{sp}=e^{\xi_0}\sqrt{m^2+l(l+1)}
= \left(\frac{m^2+l(l+1)}{m^2+l(l+3)}\right)^{1/2}{\mathcal K}_{dS}
\,.
\label{kappads}
\end{equation}
For vanishing $m$ and $l>0$, the two quantities ${\mathcal K}_{dS}$ and ${\mathcal K}_{sp}$  differ by up to 30\% and approach each other with increasing $l$. Trivially at large $l$,  but also at  small $l$ where the slope of $\omega$ as function of ${\mathcal K}$ is of the order of $10^{-3}$, the am-f relations  are  only weakly affected,\,i.e.,\,with the exception of the  $l=0,m=0$ case, the  am-f  relations are accurately described by the near horizon approximation (\ref{kappads}).

Other examples where this method for calculating the  am-f  relations  and the thermodynamic quantities can be applied to  are,
\begin{itemize}
\item    Schwarzschild metric  
\begin{equation}f(r)= 1-R_S/r,\quad \quad R_S= 2GM, 
\quad \kappa=1/2R_S\,, \quad \kappa r_0=\frac{1}{2}\,,
\label{SDM}
\end{equation}
\item Schwarzschild/AdS metric
\begin{equation}f(r)=1-\frac{R_S}{r}+\frac{r^2}{R^2}\,,\quad \kappa=\frac{1}{R\,b^2(\rho)}\big(\rho+b^3(\rho)\big)\,,\quad \kappa r_0=\frac{1}{b(\rho)}\big(\rho+b^3(\rho)\big)\,,
\label{SDAM}
\end{equation}
$$\text{with}\quad \rho=\frac{R_S}{2R}\,,\quad b(\rho) = \rho^{1/3} \Big(\big(\sqrt{1+1/27\rho^2}+1\big)^{1/3}-\big(\sqrt{1+1/27\rho^2}-1\big)^{1/3}\Big)\,,$$
\item Reissner-Nordstr\"om metric 
\begin{equation}f(r)=1-\frac{R_S}{r}+\frac{R^2}{r^2}\,,\quad R=\ell_P\,Q,\quad 
\text{Planck length}\,\ell_P,\;\;
\text{charge}\, Q,
\label{RNM}
\end{equation}
$$\rho= \sqrt{1-4R^2/R_S^2}\,,\quad \kappa=\frac{2\rho}{R_S(1+\rho)^2}\,,\quad \kappa  r_0=\frac{\rho}{1+\rho}\,.$$
\end{itemize}
In concluding this section  we emphasize the universality of the Rindler space m-f relations (\ref{bocu}).   Having determined these quantities for a sufficiently large number of modes, as shown in Fig.\,\ref{disprel},  the (discrete) eigenvalues $\omega(n,l,m,r_0,\xi_0)$ for any    static spherically symmetric space  can be read off from this figure by identifying ${\mathcal K}$ with  ${\mathcal K}_{sp}$ (Eq.\,(\ref{spmcd})) and by taking into account that the scale of any dimensionful quantity is given by the appropriate power of the surface gravity.
Evaluation of thermodynamic quantities requires summation over the angular momenta $l$ and  the number of zeroes $n$. If the distance to the horizon satisfies  $e^{\xi_0}\ll 1$ significant contributions to the sum over angular momenta can be expected  only if $l\gg 1$ and  the summation  can  be replaced by integration over $l$, cf.\,Eq.\,(\ref{spmcd}), (the summation over $n$ is not replaced by an integration),
\begin{equation*}
(2l+1) dl = r_0^2\,e^{-2\xi_0} d ({\mathcal K}^2)\,.
\end{equation*}
Therefore the Eqs.\,(\ref{omk}-\ref{o0n}) apply with the area given by,
\begin{equation}
\mathcal{A}= 4 \pi r_0^2\,,
\label{ards}
\end{equation}
while  the quantities $\zeta_n(\beta,\omega_n^0)$ (Eq.\,(\ref{zet2})) are ``universal'',\,i.e.,\,independent of the parameters of the static, spherically symmetric  metrics with a non-extremal horizon.
Qualitatively, this universality was shown in reference \cite{PADM86}. Since based on 't Hooft's approximation  \cite{tHOO85} (cf.\,Eq.\,(\ref{cond})) the expressions for the thermodynamic quantities however are incorrect.
\section{Conclusions}
Momentum- or angular momentum-frequency  relations have been shown to be the generic  tools for calculating  the kinematics of scalar fields in static space-times with a horizon. In particular,  in a large range of the kinematics, they exhibit properties which, after choosing appropriate scales, are universal,\,i.e.,\,independent of the details of the space-times, as our explicit  comparison of m-f  and am-f relations  of Rindler and de Sitter space respectively demonstrates. These  relations  provide   a direct avenue  to approximate analytical and accurate numerical computation of  the density of states, the essential ingredient for  the thermodynamics of fields in spaces with a horizon.
The central results for the thermodynamic quantities  is summarized in  expression (\ref{appd}) and  Table \ref{tdl1} which imply that up to a correction of less than 1\%, partition function and entropy are generated by a single mode,\,i.e.,\,black holes are low temperature systems. 
This property applies not only for the scalar fields in Rindler spaces but also, as we have seen explicitly, for fields in de Sitter space and more generally in static, spherically symmetric spaces. 
Our results are in conflict with most of the results obtained by applying the brick wall method. In order to arrive at closed expressions it has become common to replace the discrete spectrum of eigenmodes by a continuous one. We have shown in detail that this procedure cannot be justified and gives rise to values of the thermodynamic quantities which are too large by two orders of magnitude.

New insights  into the dynamics of quantum fields  of higher spin, in particular of photons,\,cf.\,\cite{KABT95},\,\cite{COLE98},\,\cite{DOWA12} and gravitons via  m-f or am-f relations can be expected. The imaginary parts of the corresponding stationary propagators \cite{LOY11},  closely related to the  m-f relations, indicate significant differences between  fields of different spin. Also the application  to   fields in  rotating black holes \cite{HKPS97},\,\cite{MUKO00},\,\cite{CHLY09} promises   to introduce a new element in the role of the probably complex m-f and am-f  relations. With the one  mode dominance  of the thermodynamic quantities, a hidden ``parameter'' specifying  the boundary condition, emerges which, as we have seen,\,(cf.\,Eq.\,(\ref{dineu})), influences severely partition function and entropy and needs to be determined. 

\appendix*\section{Failure of the Modified WKB approximation}
In order to identify in   detail the large effect of  replacing the  $n$-sum with $n\ge 1$   by an integral with $0$ as lower limit
we introduce the   following two functions $\sigma(\omega,\nu_0)$ and $\rho(\nu)$, cf.\,Eq.\,(\ref{cond}),
\begin{equation}
\sigma(\omega,\nu_0)=\frac{\beta}{\pi}\,\hat{\mathcal K}^2_1(\omega)\int_{\nu_0}^\infty d\nu \frac{\nu^3}{e^{\beta\nu\omega}-1}\,,\hspace{.65cm} \rho(\nu)=\frac{\beta}{\pi}\,\nu^3\int_0^\infty d\omega \frac{1}{e^{\beta\nu\omega}-1}\hat{\mathcal K}^2_1(\omega)\,,
\label{rho1}
\end{equation}
which, if integrated,  
\begin{equation*}
\ln \hat{Z}_{apx}=\frac{{\mathcal A}}{4\ell^2}\,\int_0^\infty d\omega\,\sigma(\omega,\nu_0)=\frac{{\mathcal A}}{4\ell^2}\,\int_{\nu_0}^\infty d\nu\,\rho(\nu)\,,
\end{equation*}
yield the approximate value of the partition function (\ref{wkbwr}) provided  $\nu_0=0$.
\begin{figure}[h!]
\begin{center}
\hspace{-9.cm}
\includegraphics[width=.5\linewidth]{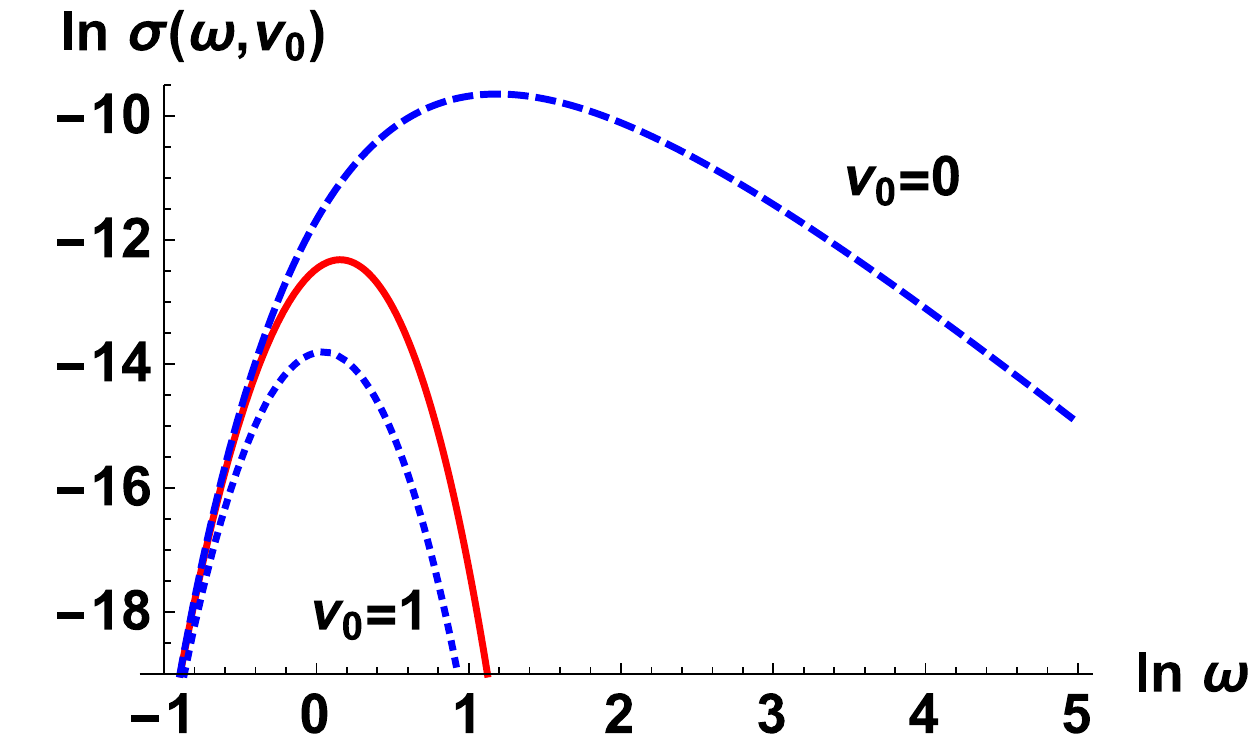}
\\
\vspace{-5.3cm}\hspace{7.9cm}
\includegraphics[width=.5\linewidth]{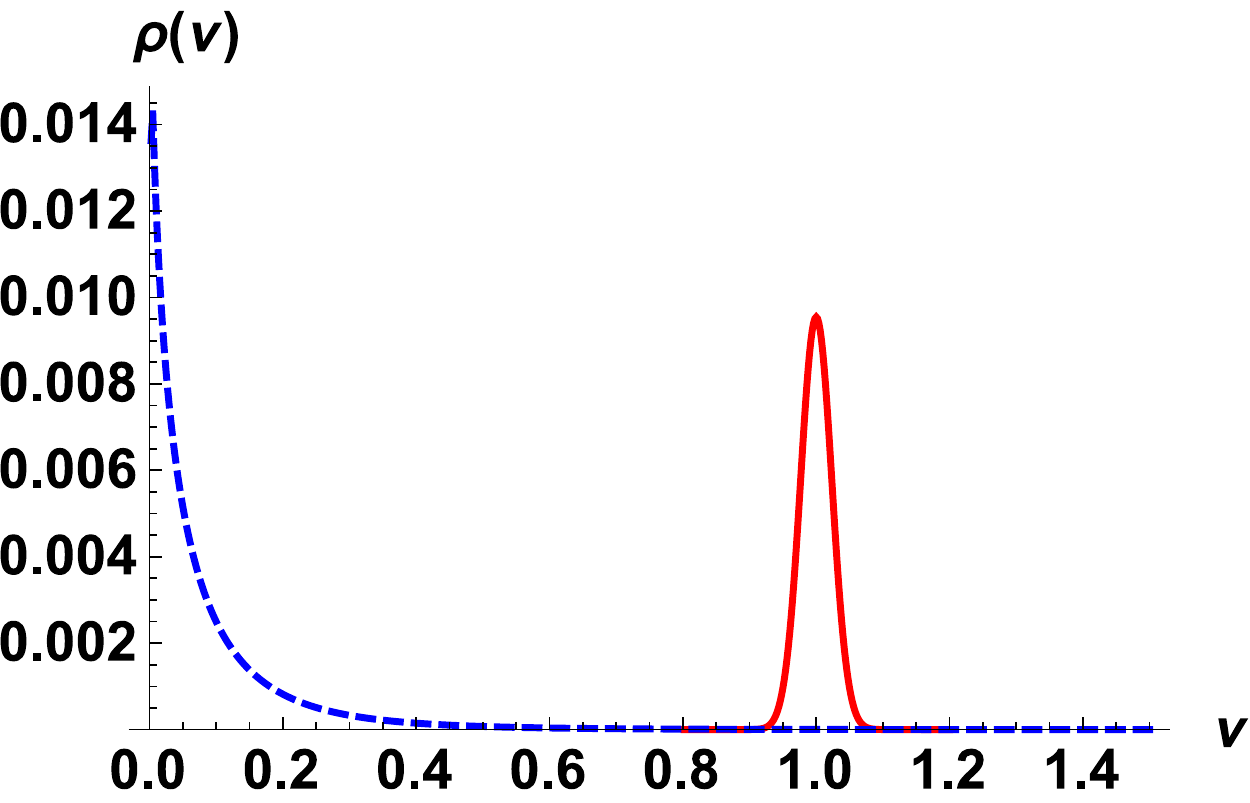}
\caption{Left: Solid (red) line:  Density $\hat{\sigma}(\omega)$ cf.\,Eq.\,(\ref{apwkb}). Dashed and dotted (blue) lines: $\sigma(\omega,\nu_0)$, Eq.\,(\ref{rho1}), with $\nu_0=0$  and  $\nu_0=1$ respectively
Right: Dotted (blue) line: density $\rho(\nu)$  (Eq.(\ref{rho1})), solid (red) line: the  $n=1$ contribution  multiplied by a factor 235  with a Gaussian distribution replacing  $\delta(\nu -1)$\,(cf.\,text).}
\end{center}
\vskip -.7cm
\label{dsin}
\end{figure}
The left hand side of Fig.\,5
demonstrates the high sensitivity of  the partition function $\hat{Z}_{apx}$ when varying  the lower limit of the $\omega$ integration.  Replacing the lower limit $\nu_0=0$ of the ``standard'' approximation (\ref{cond}) by $\nu_0=1$ reduces the value of the maximum of $\sigma(\omega,\nu)$ by a factor of 67  and  the corresponding half width   by a factor of 5. 
As the  right hand side of Fig.\,5
shows, the dominant contribution to the partition function actually arises   from values of $\nu$ in the interval  $0<\nu\le 0.4$ and the integration over  the interval $0\le\nu\le 1$ reproduces  the analytically determined value of the partition function (\ref{wkbwr}) up to 0.1\%.  

In summary,  it is not a large number of modes which contribute and give rise to a large value of the partition function which would justify the approximation (\ref{cond}).  Rather it is the contributions from the  unphysical region, $\nu<1$, which generate the large value of $\ln \hat{Z}_{apx}$ (Eq.\,(\ref{wkbwr})) exceeding the correct value by a factor of 235. 

\section*{Acknowledgments}
F.L. is grateful for the support   and  hospitality at  the En'yo Radiation Laboratory and the Hashimoto Mathematical Physics Laboratory of the  Nishina Accelerator Research Center at RIKEN. K.Y. thanks Drs. N. Iizuka, T. Noumi and N. Ogawa for useful discussions on de Sitter space. This work is supported in part by the Grant-in-Aid for Scientific Research from MEXT (No.\,22540302).    

\end{document}